\documentclass[amsmath,prl,aps,twocolumn]{revtex4}
\usepackage{amsthm,amsfonts}

\def\req#1{(\ref{#1})}
\newcommand{\D}{{\rm d}}

\begin{document}
\title{Repulsive Casimir forces}
\author{O. Kenneth, I. Klich, A. Mann and M. Revzen}%
\email{kenneth@physics.technion.ac.il; klich@tx.technion.ac.il}
\affiliation{Department of Physics,
\\ Technion - Israel Institute of Technology, Haifa 32000 Israel}
\date{February 2002}%

\begin{abstract}
We discuss repulsive Casimir forces between dielectric materials
with non trivial magnetic susceptibility. It is shown that
considerations based on naive pair-wise summation of Van der Waals
and Casimir Polder forces may not only give an incorrect estimate
of the magnitude of the total Casimir force, but even the wrong
sign of the force when materials with high dielectric and magnetic
response are involved. Indeed repulsive Casimir forces may be
found in a large range of parameters, and we suggest that the
effect may be realized in known materials. The phenomenon of
repulsive Casimir forces may be of importance both for
experimental study and for nanomachinery applications.
\end{abstract}
 \maketitle \vskip 2mm

It is well known that the fluctuations of electromagnetic fields
in vacuum or in material media depend on the boundary conditions
imposed on the fields. This dependence gives rise to forces which
are known as Casimir forces, acting on the boundaries. The best
known example for such forces is the attractive force experienced
by parallel conducting plates in vacuum \cite{Cas48}. Casimir
forces between similar, disjoint objects such as two conducting or
dielectric bodies are known in most cases to be attractive
\cite{Kenneth1} and are sometimes viewed as the macroscopic
consequence of Van der Waals and Casimir-Polder attraction between
molecules.

In view of the dominance of the Casimir forces at the nanometer
scale, where the attractive force could lead to restrictive limits
on nanodevices \cite{bordagn,buks}, the study of repulsive Casimir
forces is of increasing interest.

Repulsive Van der Waals forces are known to be possible if the
properties of the intermediate medium are intermediate between the
properties of two polarizable molecules \cite{Isra}. In such cases
the Hamaker constant becomes negative, a property which was
successfully employed to explain the wetting properties of liquid
helium \cite{Dzyalo61}. How can one get a repulsive behavior when
the intermediate substance is vacuum? A partial answer can be
obtained from the observation that a purely magnetically
polarizable particle repels a purely electrically polarizable
particle \cite{Bo74}. Motivated by this result, Boyer, following
Casimir's suggestion, studied inter plane Casimir force with one
plate a perfect conductor while the other is infinitely permeable.
He showed that in this case the plates repel \cite{Bo74}. This
problem was reconsidered since in \cite{Alves,daSilva}. However,
one must note that for most molecules the magnetic polarizability
is negligible compared to the electric polarizability. Indeed, in
many of the treatments of the subject it is assumed that the Van
der Waals interaction is dominated by the dielectric behavior of
the materials. If one then considers the pair-wise summation of
Van der Waals and Casimir-Polder forces as an approximation to the
force between materials, the result is generally attractive.

Calculations of the interaction between macroscopic bodies by
summation of pair - interactions is based on the assumption of
additivity of the interatomic interaction energies, which is only
justified within second order perturbation theory \cite{London30}.
It was pointed out by Axilrod and Teller \cite{Axilrod43} that
many-particle interactions may lead to substantial corrections to
the so called 'additive' result. Sparnaay \cite{Sparnaay59}
estimated the corrections for some simple many-body systems to be
as large as $30\%$. These corrections are usually taken to affect
the {\it magnitude} of the force but not its sign.

In this Letter we emphasize that for materials with high magnetic
susceptibility pair-wise summation  is no longer a good
approximation for the macroscopic Casimir force, due to collective
response of the material. Indeed, we show that these
considerations may not only give an incorrect estimate of the
magnitude of the force, but in some cases even of its sign! Thus
restrictions imposed on the sign of the force from pair wise
consideration can be misleading, and repulsive forces can be
expected in a wider range of dielectric-magnetic materials. We
study the Casimir force between materials with general
permittivity and permeability and show that for large permeability
and permittivity, the transition between attractive and repulsive
behavior depends only on the impedance
$Z=\sqrt{\mu\over\epsilon}$. In addition we show that at high
temperatures there is always attraction, and thus in some cases
the force changes sign as the temperature is increased.



We start by examining the pair interaction. The Casimir-Polder
potential between two polarizable particles $A$ and $B$ is given
by \cite{Berestetskii,Feinberg70}:
\begin{eqnarray}\label{pairPotential}
& U(r)=-{\hbar c\over 4 \pi r^7}[23(\alpha_E^A
\alpha_E^B+\alpha_M^A \alpha_M^B)-7(\alpha_E^A
\alpha_M^B+\alpha_M^A \alpha_E^B)]
\end{eqnarray}
where $\alpha_E,\alpha_M$ are the electric and magnetic
polarizability of the particles. From this equation it is
immediate that repulsion between particles can be obtained. For
example, a purely electrically polarizable particle will repel a
purely magnetically polarizable particle. What does this tell us
about materials described by a given permittivity and
permeability?

As an illustration for the subtle character pair wise summations
may have we consider two materials with permeability and
permittivity $\epsilon_i,\mu_i$ ($i=1,2$). When one considers two
polarizable balls in vacuum \cite{Mostepanenko} the coefficient of
the force can be read from \req{pairPotential} to be of the form
\begin{eqnarray}
& F_C\propto -
23{\epsilon_1-1\over\epsilon_1+2}{\epsilon_2-1\over\epsilon_2+2}
-23{\mu_1-1\over\mu_1+2}{\mu_2-1\over\mu_2+2}+\\ \nonumber &
7{\epsilon_1-1\over\epsilon_1+2}{\mu_2-1\over\mu_2+2}
+7{\epsilon_2-1\over\epsilon_2+2}{\mu_1-1\over\mu_1+2}\,.
\end{eqnarray}
To simplifies the last expression we set $\mu_1=1$. In this case
\begin{eqnarray}\label{FC}
& F_C\propto -{\epsilon_1-1\over\epsilon_1+2}\Big(
23{\epsilon_2-1\over\epsilon_2+2}-7{\mu_2-1\over\mu_2+2}\Big)
\end{eqnarray}
It can easily be shown that for $\epsilon_2>{37\over 16}$ the
force \req{FC} is negative for any $\mu_2$, and thus two such
balls attract. Thus if one regards the two materials to be made of
such "balls", and use as an approximation to the Casimir force
summation of pairs of these, one comes to the conclusion that
there is attraction whenever $\epsilon_2>{37\over 16}$ and
$\mu_1=1$. Next, however, looking at the whole Casimir energy we
demonstrate that this last statement is wrong.


The Casimir interaction between two polarizable materials can be
conveniently expressed in terms of the reflection coefficients at
the boundaries (\cite{Kenneth99,bordagn}). The Casimir energy per
unit area of two infinite slabs separated by a distance $a$ is
given by:
\begin{eqnarray}
& E_C={1\over 2}\int{\D^3{\bf
k}\over(2\pi)^3}\big\{\ln(1-r(\epsilon_1,\mu_1)r(\epsilon_2,\mu_2)e^{-2a|{\bf
k}|})+\\ \nonumber &
\ln(1-r(\mu_1,\epsilon_1)r(\mu_2,\epsilon_2)e^{-2a|{\bf
k}|})\big\}
\end{eqnarray}
where the two terms on the right correspond to TE and TM modes.
Here $r(\epsilon,\mu)$ for the TE mode is given by \footnote{$r$
is related to the usual reflection coefficient at the interface
between vacuum and a medium with given $\epsilon$ and $\mu$ (see
for example eq.7.39 in \cite{jackson75}) by substituting
${k_{\bot}}^2=\omega^2-{k_{\|}}^2$ and wick rotating
$\omega\rightarrow ik_t$. The angle $\theta$ in \req{reflection}
is defined by $\cos\theta={k_t\over\sqrt{{k_t}^2+{k_{\|}}^2}}$}
:
\begin{eqnarray}\label{reflection}
r(\epsilon,\mu,\theta)={\sqrt{\epsilon\mu\,\cos^2(\theta)+
\sin^2(\theta)}-\mu\over\sqrt{\epsilon\mu\,\cos^2(\theta)+
\sin^2(\theta)}+\mu}
\end{eqnarray}
and by a similar expression with $\epsilon\leftrightarrow\mu$ for
the TM mode (i.e. $r_{TM}(\epsilon,\mu)=r_{TE}(\mu,\epsilon)$).
Hence the energy is given by:
\begin{eqnarray}
& E_C={1\over8\pi^2}\int_0^{\pi} \D\theta \sin\theta\\ \nonumber &
\int_0^{\infty}\D k\,
k^2\big\{\ln(1-r(\epsilon_1,\mu_1)r(\epsilon_2,\mu_2)e^{-2ak})
+\epsilon\leftrightarrow\mu\big\}
\end{eqnarray}
The $k$ integration can be done by expanding the logarithm and
integrating term by term, to obtain
\begin{eqnarray}\label{TotalEnergy}
& E_C={-1\over 32\pi^2 a^3}\int_0^{\pi}{\rm
Li}_4[r(\epsilon_1,\mu_1,\theta)r(\epsilon_2,\mu_2,\theta)]\sin\theta\D\theta
\\ \nonumber & +\epsilon\leftrightarrow\mu\, ,
\end{eqnarray}
where ${\rm Li}_4(x)=\sum_{n=1}^{\infty}{x^n\over n^4}$ is a
polylogarithmic function.

In figure \req{F} we show positive and negative domains of the
Casimir energy \req{TotalEnergy}. Note that the condition
$\epsilon_2<{37\over 16}$ is not fulfilled in the repulsive
regime, contradicting the argument given above based on the
pair-wise attraction picture. Thus determination of the sign of
the Casimir force in the general case must involve the full
expression \req{TotalEnergy}, which predicts large repulsive
regimes.
\begin{figure}\center{\input epsf \epsfxsize=2.0in
\epsfbox{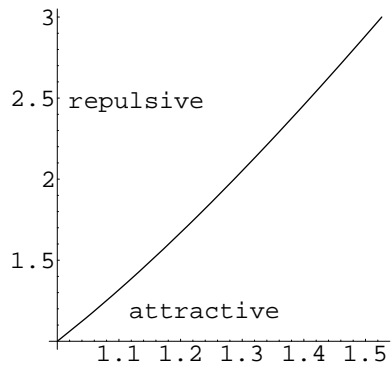}} \vskip 0.4in \center{\input epsf
\epsfxsize=2in \epsfbox{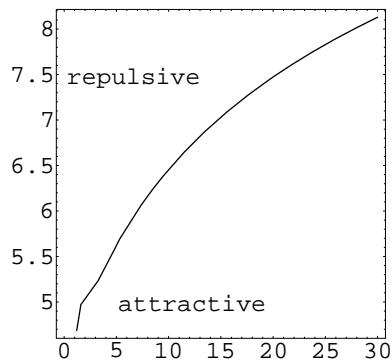}} \caption{repulsive and
attracting regions: $(a)$ In $\epsilon_2,\mu_2$ plane for
$\epsilon_1=2$ and $\mu_1=1$ and $(b)$ In $\epsilon_1,\epsilon_2$
plane $\mu_1=1$ $\mu_2=20$.} \label{F}
\end{figure}

Although the full expression for the energy \req{TotalEnergy} is
quite complicated, there is a simple statement to be made
regarding the direction of the force: The border line between
repulsive and attracting regimes for large values in the
$\epsilon_1,\mu_1$ plane (as seen in the figures), is always
linear, i.e. of the form ${\epsilon_1\over\mu_1}\rightarrow\,{\rm
const.}$ when $(\epsilon_1,\mu_1)\rightarrow\infty$. This can be
seen from the following argument: for large $\mu$ and $\epsilon$
we have
\begin{eqnarray}\label{approxr}
r(\epsilon,\mu,\theta)\sim{\sqrt{{\epsilon\over\mu}}\,|\cos(\theta)|
-1\over{\sqrt{{\epsilon\over\mu}}\,|\cos(\theta)|+1}}
\end{eqnarray}
(and a similar expression holds for $\epsilon\leftrightarrow\mu$)
Thus in this limit the Casimir energy \req{TotalEnergy} can be
written as a function of ${\epsilon\over\mu}$ and vanishes for a
certain value of this ratio. Note that one doesn't have to use
very high values of the permittivity and permeability for the
approximation \req{approxr} to be valid (see for example fig.
\ref{F} (a)), since the terms we omit are reduced by factors of
${1\over\mu}$ or $1\over\epsilon$.
\begin{figure}\center{\input epsf \epsfxsize=2.0in
\epsfbox{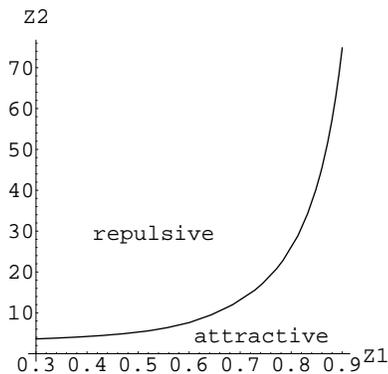}} \caption{repulsive and attracting regions in
the $Z_1,Z_2$ plane.} \label{Fgamma}
\end{figure}

We now point out several particular cases:

1. When both of the materials have high permeability and
permittivity one can use the approximation \req{approxr} for both
materials. To get an idea about the sign we approximate the
Casimir energy \req{TotalEnergy} by calculating the integral using
just the first term in the polylogarithmic function and obtain:
\begin{eqnarray}
& E_C\sim{-1\over 16\pi^2
a^3}(2+I({Z_1},{Z_2})+I({1\over{Z_1}},{1\over{Z_2}}))
\end{eqnarray}
where ${Z_i}=\sqrt{{\mu_i\over\epsilon_i}}$ are the impedances of
the materials and $I({Z_1},{Z_2})=2{{Z_2}+{Z_1}\over{Z_2}-{Z_1}}
({Z_1}\ln({1\over{Z_1}}+1)-{Z_2}\ln({1\over{Z_2}}+1))$. The border
curve defined by $E_C({Z_1},{Z_2})=0$ is shown in figure
\ref{Fgamma}.

2. If one of the bodies is a perfect conductor then for large
$\mu$ and $\epsilon$ the Casimir energy is given by:
\begin{eqnarray}
& E_C={-1\over 32\pi^2 a^3}\int_0^{\pi}{\rm
Li}_4[r(\epsilon_1,\mu_1,\theta)]+{\rm
Li}_4[-r(\mu_1,\epsilon_1,\theta)]\sin\theta\D\theta\\ \nonumber &
\sim{-1\over 16\pi^2 a^3}\Big({14\over
8{Z_1}}\ln({Z_1}+1)+{3\over8}(1-6{Z_1}\ln(1+{1\over{Z_1}}))\Big)
\end{eqnarray}
where the first two terms in the ${\rm Li}_4$ series where used.
In this case for high permeability and permittivity the transition
from attractive to repulsive regime takes place at ${Z_1}=1.037$
i.e. $\mu\sim 1.08\,\epsilon$.

3. The Casimir energy \req{TotalEnergy} can be most easily
analyzed in the uniform velocity of light (UVL) case \footnote{The
UVL condition implies $\epsilon\mu=const.$ It was introduced by T.
D. Lee \cite{Lee80} in an effective description of gluon fields
and used by Brevik and Kolbendsvet \cite{Brevik82} in the context
of electromagnetic Casimir forces.}. In this case the reflection
coefficients \req{reflection} are independent of the angle,
namely: $r(\epsilon,\mu)={1-\mu\over 1+\mu}$. This makes the
$\theta$ integral in \req{TotalEnergy} trivial, with the result:
\begin{eqnarray}
E_C={1\over 8 \pi^2a^3} {\rm Li}_4\Big(\big({1-\mu_1\over
1+\mu_1}\big)\big({1-\mu_2\over 1+\mu_2}\big)\Big)
\end{eqnarray}
This result agrees with the result obtained in \cite{Klich01} for
a dilute medium (i.e. $|\mu_i-1|<<1$ for $i=1,2$). In this case
the force becomes repulsive if $\mu_1>1$ and $\mu_2<1$ or vice
versa. However, the condition $\epsilon={1\over \mu}$ then implies
that one of the materials will have $\epsilon(\omega)<1$ on the
imaginary axis which can be shown to be inconsistent with general
properties of the dielectric function of a realistic material
\footnote{This is a simple result of the Kramers-Kr$\ddot{o}$nig
relations. Note, however that if the intermediate vacuum is to be
filled with another substance the relevant properties are relative
to the intermediate substance, and may yield repulsive behavior.}.

The leading term in the high
temperature expansion for the free energy is of the form:
\begin{equation}
F_C=-{\kappa_B T\over 16 \pi a^2}\Big({\rm Li}_3({1-\mu_1\over
1+\mu_1}{1-\mu_2\over 1+\mu_2})+{\rm Li}_3({1-\epsilon_1\over
1+\epsilon_1}{1-\epsilon_2\over 1+\epsilon_2})\Big)
\end{equation}
This expression leads to an attractive force for any values of the
permeability and permittivity provided $\mu_i,\epsilon_i>1$
($i=1,2$). Thus even if at low temperature we had repulsion the
force will change sign as we heat the system. However, the sign
change is at temperature scales of $\kappa_B T\sim{\hbar c\over
4\pi a}$ which at $100nm$ is of the order of $1000^{\circ}K$. This
phenomenon can be qualitatively explained as follows: The dominant
contribution to the free energy at high temperatures is due to
static configurations (zero modes), since contributions from other
Matsubara frequencies are exponentially suppressed. In particular
the magnetic-electric interactions are non-static by nature (there
is no static interaction between a magnetic dipole and an electric
dipole). As a result the magnetic-electric part of the interaction
which are responsible for repulsive behavior (see eq.
\req{pairPotential}) will vanish in the high temperature limit.

In view of the growing interest and possibilities of measuring the
Casimir effect \cite{bordagn,buks,Chan01,Chen02} we wish to point
out some advantages that repulsive Casimir forces might have for
actual measurements: First, there is no problem of stiction, i.e.
even for very close separations, the materials don't collapse on
each other, in contrast to the usual effect where it can sometimes
be difficult to hold them apart, a property which might be crucial
for construction of nanomachines. Moreover it might be an easier
task to align two materials in parallel, since this will be their
natural tendency (in places where the two materials get closer the
repulsion is stronger). We would also wish to note that although
it is true that for many materials the magnetic response is
negligible, there are classes of materials with high permeability,
such as ferrites and garnets (notably YIG) which may be suitable
for constructing a demonstration of repulsive Casimir forces.

We wish to thank J. Genossar for discussions. The work was
supported by the fund for promotion of research at the Technion
and by the Technion VPR fund. 
\end{document}